\documentclass[10pt]{amsart}
\usepackage{amsfonts,amssymb}
\RequirePackage{ifpdf}
\usepackage{amsmath}
\usepackage{color}
\usepackage{pstcol,pst-node}
\usepackage{graphics}
\usepackage{epic}
\usepackage{curves}

\def\appendix#1{
\addtocounter{section}{1} \setcounter{equation}{0}
\renewcommand{\thesection}{\Alph{section}}
\section*{Appendix \thesection\protect\indent\quad
#1}
}
\renewcommand{\theequation}{\thesection.\arabic{equation}}

\catcode`\@=11
\def\marginnote#1{}

\newcount\hour
\newcount\minute
\newtoks\amorpm
\hour=\time\divide\hour by60 \minute=\time{\multiply\hour by60
\global\advance\minute by-\hour}
\edef\standardtime{{\ifnum\hour<12 \global\amorpm={am}%
        \else\global\amorpm={pm}\advance\hour by-12 \fi
        \ifnum\hour=0 \hour=12 \fi
        \number\hour:\ifnum\minute<10 0\fi\number\minute\the\amorpm}}
\edef\militarytime{\number\hour:\ifnum\minute<100\fi\number\minute}

\newcommand{\tcb}{\textcolor{blue}}

\newcommand{\tcw}{\textcolor{white}}

\def\draftlabel#1{{\@bsphack\if@filesw {\let\thepage\relax
      \xdef\@gtempa{\write\@auxout{\string
          \newlabel{#1}{{\@currentlabel}{\thepage}}}}}\@gtempa \if@nobreak
    \ifvmode\nobreak\fi\fi\fi\@esphack} \gdef\@eqnlabel{#1}}
    \def\@eqnlabel{}
\def\@vacuum{}
\def\draftmarginnote#1{\marginpar{\raggedright\scriptsize\tt#1}}

\def\draft{
  \oddsidemargin -.5truein
  \def\@oddfoot{\footnotesize \sl preliminary draft \hfil
    \rm\thepage\hfil\sl\today\quad\militarytime}
  \let\@evenfoot\@oddfoot \overfullrule 3pt
    \let\label=\draftlabel
    \let\marginnote=\draftmarginnote
  \def\@eqnnum{(\theequation)\rlap{\kern\marginparsep\tt\@eqnlabel}%
    \global\let\@eqnlabel\@vacuum}

  }

\newcommand{\beq}{\begin{equation}}
\newcommand{\eeq}{\end{equation}}
\newcommand{\tr}{\,{\rm Tr}\,}
\newcommand{\ID}{1\!\!1}

\def\be{\begin{equation}}
\def\ee{\end{equation}}
\def\bea{\begin{eqnarray}}
\def\eea{\end{eqnarray}}
\def\<{\langle}
\def\>{\rangle}

\def\nn{\nonumber}

\def\Tr{{\rm Tr}}
\def\one#1{#1^{\raise5pt\hbox{$\scriptstyle\!\!\!\!1$}}\,{}}
\def\two#1{#1^{\raise5pt\hbox{$\scriptstyle\!\!\!\!2$}}\,{}}
\def\onetwo#1{#1^{\raise5pt\hbox{$\scriptstyle\!\!\!\!\!{12}$}}\,{}}
\def\otim{\mathop{\otimes}}

\def\e{e}

\newtheorem{theorem}{Theorem}[section]
\newtheorem{lm}[theorem]{Lemma}

\theoremstyle{definition}

\newtheorem{remark}[theorem]{Remark}

\newtheorem{cor}[theorem]{Corollary}
\theoremstyle{remark}

\allowdisplaybreaks

\begin{document}
\title[Shear coordinates for  $q$--$D_4$ singularity.]
{Shear coordinate description of the quantised versal unfolding of $D_4$ singularity.}
\author{Leonid Chekhov, Marta Mazzocco}

\maketitle

\begin{abstract}
In this paper by using Teichm\"uller theory of a sphere with four holes/orbifold points, we obtain a system of flat coordinates on the general affine cubic surface having a $D_4$ singularity at the origin. We show that the Goldman bracket on the geodesic functions on the four-holed/orbifold sphere coincides with the Etingof-Ginzburg Poisson bracket on the affine $D_4$ cubic. We prove that this bracket is the image under the Riemann-Hilbert map of the Poisson Lie bracket on 
$\oplus_{1}^3\mathfrak{sl}^\ast(2,{\mathbb C})$. We realise the action of the mapping class group by the action of the braid group on the geodesic functions . This action coincides with the procedure of analytic continuation of solutions of the sixth Painlev\'e equation.
Finally, we produce the explicit quantisation of the Goldman bracket on the geodesic functions on the four-holed/orbifold sphere  and of the braid group action.
\end{abstract}


\section{Introduction}
The main object studied in this paper is the following irreducible affine cubic  $\phi\in\mathbb C[u,v,w]$ having a simple  $D_4$ singularity at the origin:
$$
\tilde\phi= u^2+ v^2 + w^2- u v w + r_1 u + r_2 v + r_3 w+ r_4, 
$$
where $r_1,r_2,r_3,r_4$ are four complex parameters. It was proved in \cite{EG} that  the following formulae define a Poisson bracket on $\mathbb C[u,v,w]$:
\be\label{eq:EGintro}
\{u,v\} =\frac{\partial\widetilde\phi}{\partial w},
\qquad \{v,w\}=\frac{\partial\widetilde\phi}{\partial u},
\qquad \{w,u\}=\frac{\partial\widetilde\phi}{\partial v},
\ee
and $\widetilde\phi$ itself is a central element for this bracket, so that the quotient space
$$
M_{\widetilde\phi}:= \mathbb C[u,v,w]\slash_{\langle\widetilde\phi=0\rangle}
$$
inherits the Poisson algebra structure. Note that $M_{\widetilde\phi}$ is the manifold of the monodromy data of the sixth Painlev\'e equation \cite{jimbo}.

In this paper we define an analytic surjective map
$$
\mu:\mathbb C^3\slash_{\langle Y_1+Y_2+Y_3={\rm const.}\rangle}\to M_{\widetilde\phi},
$$
giving rise to a system of flat coordinates $Y_1,Y_2,Y_3$ for the affine irreducible cubic surface
$M_{\widetilde\phi}$.

To achieve this, we  represent $M_{\widetilde\phi}$ in terms of geodesic functions (i.e. functions of the lengths of closed geodesic curves) on a four-holed/orbifold sphere in the Poincar\'e uniformisation. These geodesic functions are merely finite Laurent polynomials of exponentials of the shear coordinates $Y_1,Y_2, Y_3$ introduced by Penner and Thurston, and they simultaneously satisfy skein relations and the Goldman Poisson relations. In particular we prove that in the case of a four-holed/orbifold sphere the Goldman bracket coincides with 
(\ref{eq:EGintro}). Despite the fact that this geometric interpretation is only valid for real valued coordinates $Y_1,Y_2,Y_3$, resulting in real $u,v,w >2$, the map $\mu$ is analytic and can be extended to any $Y_1,Y_2,Y_3\in\mathbb C$.

We present the braid-group action both on the level of geodesic functions and on the level of
shear coordinates and provide its quantum version in terms of the quantum geodesic functions.
The Poisson brackets are constants on the space of shear coordinates so their quantisation is
 straightforward and gives rise to the quantum commutation relations between the
quantum geodesic functions.

Finally we prove that the Poisson bracket  (\ref{eq:EGintro}) on the manifold of the monodromy data $M_{\widetilde\phi}$ of the sixth Painlev\'e equation is the image  under the Riemann-Hilbert map of the Poisson Lie bracket on 
$\oplus_{1}^3\mathfrak{sl}^\ast(2,{\mathbb C})$.

\vskip 2mm \noindent{\bf Acknowledgements.} The authors are grateful to
Sasha Veselov for pointing out the reference \cite{EG} and to G. Brown, D. Guzzetti, P. Rossi and
R. Vidunas  for helpful discussions. This research was
supported by the EPSRC ARF EP/D071895/1 and RA EP/F03265X/1, by the
RFBR grants 08-01-00501 and 09-01-92433-CE, by the Grant for Support
for the Scientific Schools 195.2008.1, by the Program Mathematical
Methods for Non-linear Dynamics.

\section{Geodesic algebras for a sphere with $4$ holes/orbifold points}\label{se.4-sphere}

In this section we compute the Poisson algebra of the geodesic length functions on a sphere with four holes or orbifold points.

We use the fat--graph description of the Teichm\"uller theory of
surfaces developed in \cite{Fock1}. In this section, we are going to
adapt this description to the case of sphere $\Sigma_{0,4-j,j}$ with
$4-j$ holes and $j$ orbifold points, $j=0,1,2,3,4$. The holes have
perimeters  $P_i$, $i=1,\dots,4-j$, the orbifold points correspond
to the case when the perimeters become imaginary numbers $P_l=2\pi
i/k_l$, $l=1,\dots,j$, $k_l$ being the order of the orbifold point
(the particular case of orbifold points of order $2$ was treated in
details in~\cite{Ch1a,ChMa}).

Let us start with the case $j=0$, i.e. a sphere $\Sigma_{0,4}$ with $4$ holes and no orbifold points. A fat graph associated to a Riemann
surface with holes~\cite{Fock1}~\cite{Fock2} is a
spine $\Gamma_{g,s}$, which is a connected three-valent
graph drawn without self-intersections on $\Sigma_{g,s}$
with a prescribed cyclic ordering
of labelled edges entering each vertex; it must be a maximal graph in
the sense that its complement on the Riemann surface is a set of
disjoint polygons (faces), each polygon containing exactly one hole
(and becoming simply connected after gluing this hole).

By the Poincar\'e uniformisation theorem, the Riemann surface $\Sigma_{g,s}$ of genus $g$
and with $s$ holes points can be obtained as
$$
\Sigma_{g,s}\sim {\mathbb H}\slash \Delta_{g,s},
$$
where
$$
 \Delta_{g,s}=\langle\gamma_1\dots,\gamma_{2g+s-1}\rangle,\qquad
 \gamma_1\dots,\gamma_{2g+s-1}\in{\mathbb P}SL(2,{\mathbb R})
$$
is a Fuchsian group containing only hyperbolic elements,  the fundamental group of the
surface~$\Sigma_{g,s}$.

In the Thurston shear-coordinate description of the
Teichm\"uller spaces of Riemann surfaces with holes \cite{Fock1}, we
decompose each hyperbolic matrix $\gamma\in \Delta_{g,s}$ into a
product of the form
\begin{equation}\label{eq:decomp}
\gamma= (-1)^K R^{k_{i_p}} X_{Z_{i_p}} \dots R^{k_{i_1}}
X_{Z_{i_1}},\qquad i_j\in I,\quad k_{i_j}=1,2,\quad K:=\sum_{j=1}^p
k_{i_j}
\end{equation}
where $I$ is a set of integer indices and the matrices $R,\, L$ and
$X_{Z_i}$, $Z_i\in\mathbb R$, are defined as follows:
\begin{eqnarray}\nn
&&
R:=\left(\begin{array}{cc}1&1\\-1&0\\
\end{array}\right), \qquad
L=-R^2:=\left(\begin{array}{cc}0&1\\-1&-1\\
\end{array}\right), \\
&&
X_{Z_i}:=\left(\begin{array}{cc}0&-\exp\left({\frac{Z_i}{2}}\right)\\
\exp\left(-{\frac{Z_i}{2}}\right)&0\end{array}\right).
\label{eq:generators}
\end{eqnarray}

The set of closed geodesics on a Riemann surface $\Sigma_{g,s}$ is in the one-to-one
correspondence with conjugacy classes of elements of the Fuchsian group
$\Delta_{g,s}$ with the lengths $\ell_\gamma$ of these geodesics to be determined as
$$
e^{\ell_\gamma/2}+e^{-\ell_\gamma/2}=\tr \gamma,
$$
where we take a trace of the matrix product (\ref{eq:decomp}). We call the combination
$e^{\ell_\gamma/2}+e^{-\ell_\gamma/2}$ the {\em geodesic function} $G_\gamma$.

The fat graph for $\Sigma_{0,4}$ has the form
of the three-petal graph depicted in figure (\ref{loopinvert}) where
we also present the geodesic line corresponding to the element
$G_{1,2}$:
\be
\label{loopinvert}
{\psset{unit=0.5}
\begin{pspicture}(-5,-5)(7,5)
\newcommand{\PATTERN}[1]{%
\pcline[linewidth=1pt](0.3,0.5)(2,0.5)
\pcline[linewidth=1pt](0.3,-0.5)(2,-0.5)
\psbezier[linewidth=1pt](2,0.5)(3,2)(5,2)(5,0)
\psbezier[linewidth=1pt](2,-0.5)(3,-2)(5,-2)(5,0)
\psbezier[linewidth=1pt](2.8,0)(3.6,0.8)(4,0.7)(4,0)
\psbezier[linewidth=1pt](2.8,0)(3.6,-0.8)(4,-0.7)(4,0)
\rput(1,1.2){\makebox(0,0){$Y_{#1}$}}
\rput(4.5,2){\makebox(0,0){$P_{#1}$}}
}
\rput(0,0){\PATTERN{1}}
\rput{120}(0,0){\PATTERN{2}}
\rput{240}(0,0){\PATTERN{3}}
\newcommand{\CURVE}{%
\pcline[linecolor=red, linestyle=dashed, linewidth=1.5pt](0.1,.2)(2.1,.2)
\pcline[linecolor=red, linestyle=dashed, linewidth=1.5pt](0.1,-.2)(2.1,-.2)
\psbezier[linecolor=red, linestyle=dashed, linewidth=1.5pt](2.1,.2)(3,1.7)(4.7,1.8)(4.7,0)
\psbezier[linecolor=red, linestyle=dashed, linewidth=1.5pt]{->}(2.1,-.2)(3,-1.6)(4.7,-1.8)(4.7,0)
}
\rput(0,0){\CURVE}
\rput{120}(0,0){\CURVE}
\psarc[linecolor=red, linestyle=dashed, linewidth=1.5pt](0.1,.2){.4}{210}{270}
\end{pspicture}
}
\ee

The  algebras of geodesic length functions were constructed in
\cite{Ch1} by  postulating the Poisson relations on the level of the
shear coordinates $Z_\alpha$ of the Teichm\"uller space:
\begin{equation}
\label{eq:Poisson} \bigl\{f({\mathbf Z}),g({\mathbf
Z})\bigr\}=\sum_{{\hbox{\small 3-valent} \atop \hbox{\small vertices
$\alpha=1$} }}^{4g+2s+n-4} \,\sum_{i=1}^{3\!\!\mod 3}
\left(\frac{\partial f}{\partial Z_{\alpha_i}} \frac{\partial
g}{\partial Z_{\alpha_{i+1}}} - \frac{\partial g}{\partial
Z_{\alpha_i}} \frac{\partial f}{\partial z_{\alpha_{i+1}}}\right),
\end{equation}
where the sum ranges all the (three-valent) vertices of a graph and
$\alpha_i$ are the labels of the cyclically (counterclockwise)
ordered ($\alpha_4\equiv \alpha_1 $) edges incident to the vertex
with the label $\alpha$. This bracket gives rise to the {\it Goldman
bracket} on the space of geodesic length functions \cite{Gold}. Note that we label the six shear coordinates $Z_\alpha$ by $Y_1,Y_2,Y_3,P_1,P_2,P_3$.

In the case of $\Sigma_{0,4}$, i.e. the sphere with four holes, we consider the following generators
for the Fuchsian group $\Delta_{0,4}$
\bea
&&\gamma_1=X_{Y_1}RX_{P_1}RX_{Y_1}=\left(\begin{array}{cc}0&-e^{Y_1+\frac{P_1}2}\\e^{-Y_1-\frac{P_1}2}&-e^{-{P_1}/2}-e^{{P_1}/2}\\
\end{array}\right)\nonumber\\
&&\gamma_2=-RX_{Y_2}RX_{P_2}RX_{Y_2}L=-R\left(\begin{array}{cc}0&-e^{Y_2+\frac{P_2}2}\\e^{-Y_2-\frac{P_2}2}&-e^{-{P_2}/2}-e^{{P_2}/2}\\
\end{array}\right)L\label{M}\\
&&\gamma_3=-LX_{Y_3}RX_{P_3}RX_{Y_3}R=-L\left(\begin{array}{cc}0&-e^{Y_3+\frac{P_3}2}\\e^{-Y_3-\frac{P_3}2}&-e^{-{P_3}/2}-e^{{P_3}/2}.\\
\end{array}\right)R\nonumber,
\eea

\begin{theorem}\label{th:geo-p}
The Poisson algebra of geodesic  functions on the sphere with four holes is generated by
the three elements $G_{1,2}$, $G_{2,3}$ and $G_{1,3}$:
$$
G_{i,j}:=-\Tr(\gamma_i \gamma_j),\quad i<j, i,j=1,2,3,
$$
which correspond to closed paths encircling two holes without self
intersections as shown in  figure (\ref{loopinvert}). Their Poisson
brackets are given by the formulae
\begin{eqnarray}
&&
\left\{G_{1,2},G_{2,3}\right\}= G_{1,2} G_{2,3} -2 G_{1,3} -  \omega_{1,3},\label{eq:gen-poisson1}\\
&&
\left\{G_{2,3},G_{1,3}\right\}= G_{2,3} G_{1,3} -2 G_{1,2} -  \omega_{1,2},\label{eq:gen-poisson2}\\
&&
\left\{G_{1,3},G_{1,2}\right\}=  G_{1,2} G_{1,3} -2 G_{2,3} - \omega_{2,3},\label{eq:gen-poisson3}
\end{eqnarray}
where
$$
\omega_{ij}=G_i G_j+G_k G_\infty\hbox{ for  }k\neq i,j,\quad G_i=-\Tr(\gamma_i),
\quad G_\infty=-\Tr(\gamma_1\gamma_2\gamma_3).
$$
The formulae (\ref{eq:gen-poisson1}), (\ref{eq:gen-poisson2}), (\ref{eq:gen-poisson3}) define an abstract
Poisson algebra satisfying the Jacobi relations for any choice of the constants $\omega_{i,j}$. The central element of this algebra is
\beq
{\mathcal C}=G_{1,2}^2+G_{2,3}^2+G_{1,3}^2-G_{1,2}G_{2,3}G_{1,3}+G_{1,2}\omega_{1,2}
+G_{2,3}\omega_{2,3}+G_{1,3}\omega_{1,3}.
\label{M-el}
\eeq
\end{theorem}

\proof
For  convenience, we perform the change of variable ${\widetilde Y}_i= Y_i-P_i/2$, $i=1,2,3$,
after which the matrix combination $X_{{\widetilde Y}_i}RX_{P_i}RX_{{\widetilde Y}_i}$, which is the main building block in (\ref{M}) becomes merely
\begin{equation}
\label{block}
\left(\begin{array}{cc}0&-e^{\widetilde Y_i}\\e^{-\widetilde Y_i}&-G_i\\ \end{array}\right),
\end{equation}
where $G_i=e^{{P_i}/2}+e^{-{P_i}/2}$
is the trace of the monodromy around the hole. Note that thanks to the shape of the fat--graph,
$$
\{Y_i,P_j\}=0 \quad\hbox{for}\, i\neq j,
$$
so that
$$
\{\widetilde Y_i,\widetilde Y_j\}=\{Y_i,Y_j\}.
$$
The explicit form of $G_{i,j}$ are then
\begin{eqnarray}
\label{G12}
&&
G_{1,2}=e^{\widetilde Y_1+\widetilde Y_2}+
e^{-\widetilde Y_1-\widetilde Y_2}+e^{-\widetilde Y_1+\widetilde Y_2}+
G_1e^{\widetilde Y_2}+G_2e^{-\widetilde Y_1}\nn\\
&&
G_{2,3}=e^{\widetilde Y_2+\widetilde Y_3}+
e^{-\widetilde Y_2-\widetilde Y_3}+e^{-\widetilde Y_2+\widetilde Y_3}+
G_2e^{\widetilde Y_3}+G_3 e^{-\widetilde Y_2}\\
&&
G_{3,1}=e^{\widetilde Y_3+\widetilde Y_1}+
e^{-\widetilde Y_3-\widetilde Y_1}+e^{-\widetilde Y_3+\widetilde Y_1}+
G_3e^{\widetilde Y_1}+G_1e^{-\widetilde Y_3}.\nn
\end{eqnarray}

The Poisson brackets  (\ref{eq:gen-poisson1}),  (\ref{eq:gen-poisson2}) and (\ref{eq:gen-poisson3}) can be proved now by brute
force computation by applying the Poisson brackets (\ref{eq:Poisson}) \endproof

\begin{remark}
As we mentioned in the above Theorem the formulae  (\ref{eq:gen-poisson1}), (\ref{eq:gen-poisson2}), (\ref{eq:gen-poisson3}) define an abstract
Poisson algebra, i.e. we can think of $G_{i,j}$ as abstract quantities. If we impose the parametrisation (\ref{G12}), then it is straightforward to prove that the
central element $\mathcal C$ satisfies the following relation
originally due to Fricke \cite{Mag}
\begin{equation}\label{eq:fr}
\mathcal C= 4-G_1G_2G_3G_\infty-G_1^2-G_2^2-G_3^2-G_\infty^2.
\end{equation}
This relation plays a fundamental role in the theory of the sixth Painlev\'e equation which will be discussed in Section \ref{se:pain} below.
\end{remark}

\subsection{Sphere with orbifold points}\label{se:orb}

Let us consider the case  in which one or more holes in the sphere
$\Sigma_{0,4}$ are substituted by orbifold points. As mentioned
earlier, this corresponds to allowing the perimeters to become
imaginary numbers $P_l=2\pi i/k_l$, $k_l$ being the order of the
orbifold point. The Poincar\'e uniformisation theorem still holds:
$$
\Sigma_{0,4-j,j}\sim {\mathbb H}\slash \Delta_{0,4-j,j},
$$
where the Fuchsian group $\Delta_{0,4-j,j}$ is now generated by
$4-j$ hyperbolic elements and by $j$ elliptic elements satisfying
one relation. The hyperbolic elements are expressed in terms of
Thurston shear-coordinate as in (\ref{eq:decomp}) above, while the
elliptic elements are decomposed as follows: we take $G:=2\cos
(2\pi/k)$ and set the matrix
$$
F^{(R)}=\left(\begin{array}{cc}G&1\\-1&0\\ \end{array}\right)
$$
every time we go around the orbifold point counterclockwise and the matrix
$$
F^{(L)}=\left(\begin{array}{cc}0&1\\-1&-G\\ \end{array}\right)
$$
every time we go around it clockwise. When all holes are replaced by orbifold points, the generators of the Fuchsian group become
\bea
&&\gamma_1=X_{Y_1}F^{(R)}_{1}X_{Y_1}
\nonumber\\
&&\gamma_2=-RX_{Y_2}F^{(R)}_{2}X_{Y_2}L
\label{M1}\\
&&\gamma_3=-LX_{Y_3}F^{(R)}_{3}X_{Y_3}R.
\nonumber
\eea
We see that
the matrix combination $X_{Y_i}F^{(R)}_{i}X_{Y_i}$ has exactly the
form (\ref{block}) in which now  $\widetilde Y_i=Y_i$ and $G_i=2\cos
(2\pi/k_i)$, where $k_i$ is the order of the corresponding orbifold
point. We can therefore treat in a uniform way both the case of a
hole and of an orbifold point. In particular the quantities
$G_{i,j}:=-\Tr(\gamma_i \gamma_j)$, $i<j, i,j=1,2,3$ have now the
same form (\ref{G12}) with $\widetilde Y_i=Y_i$ and the parameter
$G_i=2\cos (2\pi/k_i)$, $2>G_i\ge0$:
\begin{eqnarray}
\label{G12orbifold}
&&
G_{1,2}=e^{ Y _1+ Y _2}+
e^{- Y _1- Y _2}+e^{- Y _1+ Y _2}+
G_1e^{ Y _2}+G_2e^{- Y _1}\nn\\
&&
G_{2,3}=e^{ Y _2+ Y _3}+
e^{- Y _2- Y _3}+e^{- Y _2+ Y _3}+
G_2e^{ Y _3}+G_3 e^{- Y _2}\\
&&
G_{3,1}=e^{ Y _3+ Y _1}+
e^{- Y _3- Y _1}+e^{- Y _3+Y_1}+
G_3e^{Y_1}+G_1e^{- Y_3}.\nn
\end{eqnarray}
As a result the following corollary of Theorem \ref{th:geo-p} holds true:

\begin{cor}
The Poisson algebra of geodesic length functions on the sphere with
$4-j$ holes and $j$ orbifold points, $j=1,2,3,4$,  is generated by
the three elements $G_{1,2}$, $G_{1,3}$, and $G_{2,3}$
$$
G_{i,j}:=-\Tr(\gamma_i \gamma_j),\quad i<j, i,j=1,2,3.
$$
Their Poisson brackets are given by the formulae (\ref{eq:gen-poisson1}),  (\ref{eq:gen-poisson2}) and (\ref{eq:gen-poisson3}).
\end{cor}

\proof As explained above, this is a straightforward consequence of
Theorem \ref{th:geo-p}. We present here an alternative proof  that
follows from evaluating the Poisson brackets between, say, $G_{1,2}$
and $G_{2,3}$ using the Goldman bracket \cite{Gold} and the skein
relation. For this, we introduce a new geodesic function
\be
\label{tilde-G13} {\widetilde G}_{1,3}:=+\Tr(\gamma_1\gamma_2\gamma_3\gamma_2^{-1}),
\ee
which
corresponds to the geodesic that goes around the holes/orbifold
points with the numbers 1 and 3 and goes twice around the
hole/orbifold point with the number 2. It is then easy to see that
$$
\{G_{1,2},G_{1,3}\}={\widetilde G}_{1,3}-G_{1,3}
$$
and we can use the skein relation for the product of $G_{1,2}$ and $G_{1,3}$:
$$
G_{1,2}G_{1,3}={\widetilde G}_{1,3}+G_{1,3}+G_1G_3+G_2G_\infty,
$$
where $G_\infty=-\Tr \gamma_1\gamma_2\gamma_3=e^{Y_1+Y_2+Y_3}+e^{-Y_1-Y_2-Y_3}$ is the central element
corresponding to the geodesic that goes around the last, fourth, hole. Expressing
${\widetilde G}_{1,3}$ from this relation, we immediately come to (\ref{eq:gen-poisson1})
in which we set $\omega_{1,3}:=G_1G_3+G_2G_\infty$. \endproof

\subsection{Braid group action on $\Sigma_{0,4-j,j}$}

The action of the braid-group element $\beta_{i,i+1}$, $i=1,2$, in terms of the geodesic functions corresponds to interchanging  $i$th and $(i+1)$th holes/orbifold points
resulting in a continuous deformation of loops on the four-holed sphere. On the level of the
Teichm\"uller space coordinates $Y_i$, we achieve this permutation by flipping edges. 

Here we illustrate the action of $\beta_{1,2}$: first, the one with the
label $Y_2$, second, the one with the label $Y_2+P_2$:
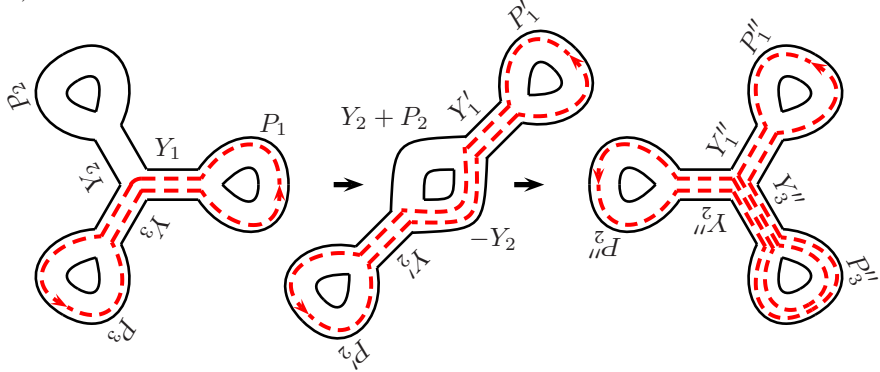
\be
\label{flips}
{\psset{unit=0.4}
\begin{pspicture}(-5,-6)(5,6)
\newcommand{\PATTERN}[1]{%
\pcline[linewidth=1pt](0.3,0.5)(2,0.5)
\pcline[linewidth=1pt](0.3,-0.5)(2,-0.5)
\psbezier[linewidth=1pt](2,0.5)(3,2)(5,2)(5,0)
\psbezier[linewidth=1pt](2,-0.5)(3,-2)(5,-2)(5,0)
\psbezier[linewidth=1pt](2.8,0)(3.6,0.8)(4,0.7)(4,0)
\psbezier[linewidth=1pt](2.8,0)(3.6,-0.8)(4,-0.7)(4,0)
\rput(1,1.2){\makebox(0,0){$Y_{#1}$}}
\rput(4.5,2){\makebox(0,0){$P_{#1}$}}
}
\rput(0,0){\PATTERN{1}}
\rput{120}(0,0){\PATTERN{2}}
\rput{240}(0,0){\PATTERN{3}}
\newcommand{\CURVE}{%
\pcline[linecolor=red, linestyle=dashed, linewidth=1.5pt](0.1,.2)(2.1,.2)
\pcline[linecolor=red, linestyle=dashed, linewidth=1.5pt](0.1,-.2)(2.1,-.2)
\psbezier[linecolor=red, linestyle=dashed, linewidth=1.5pt](2.1,.2)(3,1.7)(4.7,1.8)(4.7,0)
\psbezier[linecolor=red, linestyle=dashed, linewidth=1.5pt]{->}(2.1,-.2)(3,-1.6)(4.7,-1.8)(4.7,0)
}
\rput(0,0){\CURVE}
\rput{240}(0,0){\CURVE}
\psarc[linecolor=red, linestyle=dashed, linewidth=1.5pt](0.1,-.2){.4}{90}{150}
\end{pspicture}
\begin{pspicture}(-5,-6)(5,6)
\newcommand{\PATTERN}[1]{%
\pcline[linewidth=1pt](0.3,0.5)(2,0.5)
\pcline[linewidth=1pt](0.3,-0.5)(2,-0.5)
\psbezier[linewidth=1pt](2,0.5)(3,2)(5,2)(5,0)
\psbezier[linewidth=1pt](2,-0.5)(3,-2)(5,-2)(5,0)
\psbezier[linewidth=1pt](2.8,0)(3.6,0.8)(4,0.7)(4,0)
\psbezier[linewidth=1pt](2.8,0)(3.6,-0.8)(4,-0.7)(4,0)
\rput(1,1.2){\makebox(0,0){$Y'_{#1}$}}
\rput(4.5,2){\makebox(0,0){$P'_{#1}$}}
}
\rput{45}(1,1){\PATTERN{1}}
\rput{225}(-1,-1){\PATTERN{2}}
\psbezier[linewidth=1pt](.9,1.6)(-1.5,1.5)(-1.5,1.5)(-1.6,-.9)
\psbezier[linewidth=1pt](1.6,.9)(1.5,-1.5)(1.5,-1.5)(-.9,-1.6)
\psbezier[linewidth=1pt](.5,.5)(-.5,.5)(-.5,.5)(-.5,-.5)
\psbezier[linewidth=1pt](.5,.5)(.5,-.5)(.5,-.5)(-.5,-.5)
\rput(1.8,-1.8){\makebox(0,0){$-Y_2$}}
\rput(-1.8,2.2){\makebox(0,0){$Y_2+P_2$}}
\pcline[linewidth=2pt]{->}(-3.5,0)(-2.5,0)
\pcline[linewidth=2pt]{->}(2.5,0)(3.5,0)
\newcommand{\CURVE}{%
\pcline[linecolor=red, linestyle=dashed, linewidth=1.5pt](0.1,.2)(2.1,.2)
\pcline[linecolor=red, linestyle=dashed, linewidth=1.5pt](0.1,-.2)(2.1,-.2)
\psbezier[linecolor=red, linestyle=dashed, linewidth=1.5pt](2.1,.2)(3,1.7)(4.7,1.8)(4.7,0)
\psbezier[linecolor=red, linestyle=dashed, linewidth=1.5pt]{->}(2.1,-.2)(3,-1.6)(4.7,-1.8)(4.7,0)
}
\rput{45}(1,1){\CURVE}
\rput{225}(-1,-1){\CURVE}
\psbezier[linecolor=red, linestyle=dashed, linewidth=1.5pt](-.9,-1.25)(1.3,-1.3)(1.3,-1.3)(1.25,.9)
\psbezier[linecolor=red, linestyle=dashed, linewidth=1.5pt](-1.25,-.85)(1,-1)(1,-1)(.85,1.25)
\end{pspicture}
\begin{pspicture}(-5,-6)(5,6)
\newcommand{\PATTERN}[1]{%
\pcline[linewidth=1pt](0.3,0.5)(2,0.5)
\pcline[linewidth=1pt](0.3,-0.5)(2,-0.5)
\psbezier[linewidth=1pt](2,0.5)(3,2)(5,2)(5,0)
\psbezier[linewidth=1pt](2,-0.5)(3,-2)(5,-2)(5,0)
\psbezier[linewidth=1pt](2.8,0)(3.6,0.8)(4,0.7)(4,0)
\psbezier[linewidth=1pt](2.8,0)(3.6,-0.8)(4,-0.7)(4,0)
\rput(1,1.2){\makebox(0,0){$Y''_{#1}$}}
\rput(4.5,2){\makebox(0,0){$P''_{#1}$}}
}
\rput{60}(0,0){\PATTERN{1}}
\rput{180}(0,0){\PATTERN{2}}
\rput{300}(0,0){\PATTERN{3}}
\newcommand{\CURVE}{%
\pcline[linecolor=red, linestyle=dashed, linewidth=1.5pt](0.25,.2)(2.1,.2)
\pcline[linecolor=red, linestyle=dashed, linewidth=1.5pt](0.25,-.2)(2.1,-.2)
\psbezier[linecolor=red, linestyle=dashed, linewidth=1.5pt](2.1,.2)(3,1.7)(4.7,1.8)(4.7,0)
\psbezier[linecolor=red, linestyle=dashed, linewidth=1.5pt]{->}(2.1,-.2)(3,-1.6)(4.7,-1.8)(4.7,0)
}
\rput{60}(0,0){\CURVE}
\rput{180}(0,0){\CURVE}
\newcommand{\CURVEONE}{%
\pcline[linecolor=red, linestyle=dashed, linewidth=1.5pt](0.1,0.3)(2.2,0.3)
\pcline[linecolor=red, linestyle=dashed, linewidth=1.5pt](0.1,-0.3)(2.2,-0.3)
\psbezier[linecolor=red, linestyle=dashed, linewidth=1.5pt](2.2,0.3)(3.2,1.8)(4.8,1.8)(4.8,0)
\psbezier[linecolor=red, linestyle=dashed, linewidth=1.5pt](2.2,-0.3)(3.2,-1.8)(4.8,-1.8)(4.8,0)
\pcline[linecolor=red, linestyle=dashed, linewidth=1.5pt](-0.3,0.1)(2.4,0.1)
\pcline[linecolor=red, linestyle=dashed, linewidth=1.5pt](-0.3,-0.1)(2.4,-0.1)
\psbezier[linecolor=red, linestyle=dashed, linewidth=1.5pt](2.4,0.1)(3.2,1.4)(4.6,1.3)(4.6,0)
\psbezier[linecolor=red, linestyle=dashed, linewidth=1.5pt](2.4,-0.1)(3.2,-1.4)(4.6,-1.3)(4.6,0)
}
\rput{300}(0,0){\CURVEONE}
\end{pspicture}
}
\ee
In this picture, we also indicate the (continuous) transformation of $G_{1,3}$ that leaves it invariant (in the
new variables $Y''_i$, $P''_i$).

The resulting transformation (in terms of shifted variables $Y_i$) reads:
\bea
&Y''_1=Y_1+\log(1+G_2e^{Y_2}+e^{2Y_2}),&\quad P''_1=P_1,\nonumber\\
&Y''_2=Y_3-\log(1+G_2e^{-Y_2}+e^{-2Y_2}),&\quad P''_2=P_3,\nonumber\\
&Y''_3=-Y_2,&\quad P''_3=P_2,\nonumber
\eea
and it produces the following  formulae for the corresponding transformations of the geodesic functions:
\be
\label{beta1}
\begin{array}{ll}
\beta_{1,2}G_{1,2}=G_{1,2}, & \beta_{1,2}\omega_{1,2}=\omega_{1,2},\\
\beta_{1,2}G_{2,3}=G_{1,3}, & \beta_{1,2}\omega_{2,3}=\omega_{1,3},\\
\beta_{1,2}G_{1,3}=G_{1,2}G_{1,3}-G_{2,3}-
\omega_{2,3}, & \beta_{1,2}\omega_{1,3}=\omega_{2,3},
\end{array}
\ee
and by the same procedure:
\be
\label{beta2}
\begin{array}{ll}
\beta_{2,3}G_{1,2}=G_{2,3}G_{1,2}-G_{1,3}-\omega_{1,3} ,
& \beta_{2,3}\omega_{1,2}=\omega_{1,3},\\
\beta_{2,3}G_{2,3}=G_{2,3}, & \beta_{2,3}\omega_{2,3}=\omega_{2,3},\\
\beta_{2,3}G_{1,3}=G_{1,2}, & \beta_{2,3}\omega_{1,3}=\omega_{1,2}.
\end{array}
\ee

\begin{lm}
The transformations (\ref{beta1}), (\ref{beta2}) satisfy the braid-group relations and the element (\ref{M-el})
is invariant w.r.t. the braid-group transformations.
\end{lm}

\proof This result is proved by straightforward computations, and in the context of the Painlev\'e sixth equation was proved in  \cite{iwa}.\endproof

\section{Quantised Poisson algebra}\label{se:quantum}
In the quantum version, we introduce the Hermitian operators $Y^\hbar_i$ subject to the commutation
inherited from the Poisson bracket of $Y_i$:
$$
[Y^\hbar_i,Y^\hbar_{i+1}]=i\pi \hbar \{Y_i,Y_{i+1}\}=i\pi \hbar,\quad i=1,2,3,\ i+3\equiv i.
$$
Observe that thanks to this fact, the commutators $[Y^\hbar_i,Y^\hbar_{j}]$ are always numbers and therefore we have
$$
\exp\left({a Y_i^\hbar}\right) \exp\left({b Y_j^\hbar}\right) =
\exp\left(a {Y_i^\hbar}+b {Y_i^\hbar}+\frac{ab}{2}[Y^\hbar_i,Y^\hbar_{j}]\right) ,
$$
for any two constants $a,b$. Therefore we have the Weyl ordering:
$$
e^{Y^\hbar_{1}+Y^\hbar_{2}}=q^{\frac{1}{2}}e^{Y^\hbar_{1}}e^{Y^\hbar_{2}}=q^{-\frac{1}{2}}e^{Y^\hbar_{2}}e^{Y^\hbar_{1}},\quad q\equiv e^{-i\pi\hbar}.
$$
After quantisation, the central elements $G_1,G_2,G_3$ remain
central and non--deformed, so we preserve the previous notation for
them. We assume that the expression for $G^\hbar_{1,2},
G^\hbar_{2,3}, G^\hbar_{1,3}, $ have precisely the form of
(\ref{G12}) or (\ref{G12orbifold})  with $Y^\hbar_{i}$ substituted
for the respective $Y_{i}$, in order to ensure the Hermiticity of
$G^\hbar_{i,j}$: \
$\bigl[G^\hbar_{i,j}\bigr]^\dagger=G^\hbar_{i,j}$.

We have the corresponding deformations of the Poisson relations, which become commutation relations between $G^\hbar_{i,j}$:
\bea
q^{-1/2}G^\hbar_{1,2}G^\hbar_{2,3}-q^{1/2}G^\hbar_{2,3}G^\hbar_{1,2}&=&(q^{-1}-q)G^\hbar_{1,3}+(q^{-1/2}-q^{1/2})\omega_{1,3}\nn\\
q^{-1/2}G^\hbar_{2,3}G^\hbar_{1,3}-q^{1/2}G^\hbar_{1,3}G^\hbar_{2,3}&=&(q^{-1}-q)G^\hbar_{1,2}+(q^{-1/2}-q^{1/2})\omega_{1,2}
\label{q-comm}\\
q^{-1/2}G^\hbar_{1,3}G^\hbar_{1,2}-q^{1/2}G^\hbar_{1,2}G^\hbar_{1,3}&=&(q^{-1}-q)G^\hbar_{2,3}+(q^{-1/2}-q^{1/2})\omega_{2,3}\nn
\eea
The action of the quantum braid group is given by:
\be
\label{beta12-q}
\begin{array}{ll}
\beta_{1,2}G^\hbar_{1,2}=G^\hbar_{1,2} & \beta_{1,2}\omega_{1,2}=\omega_{1,2}\\
\beta_{1,2}G^\hbar_{2,3}=G^\hbar_{1,3} & \beta_{1,2}\omega_{2,3}=\omega_{1,3}\\
\beta_{1,2}G^\hbar_{1,3}={\widetilde G}^\hbar_{2,3}=q^{1/2}G^\hbar_{1,2}G^\hbar_{1,3}-q G^\hbar_{2,3}-q^{1/2}\omega_{2,3}
& \beta_{1,2}\omega_{1,3}=\omega_{2,3}\\
\tcw{\beta_{1,2}G^\hbar_{1,3}={\widetilde G}^\hbar_{2,3}}=q^{-1/2}G^\hbar_{1,3}G^\hbar_{1,2}-q^{-1}G^\hbar_{2,3}-q^{-1/2}\omega_{2,3} &
\end{array}
\ee
\be
\label{beta23-q}
\begin{array}{ll}
\beta_{2,3}G^\hbar_{1,2}={\widetilde G}^\hbar_{1,3}=q^{1/2}G^\hbar_{2,3}G^\hbar_{1,2}-q G^\hbar_{1,3}-q^{-1/2}\omega_{1,3}
& \beta_{2,3}\omega_{1,2}=\omega_{1,3}\\
\tcw{\beta_{2,3}G^\hbar_{1,2}={\widetilde G}^\hbar_{1,3}}=q^{-1/2}G^\hbar_{1,2}G^\hbar_{2,3}-q^{-1}G^\hbar_{1,3}-q^{-1/2}\omega_{1,3}&\\
\beta_{2,3}G^\hbar_{2,3}=G^\hbar_{2,3} & \beta_{2,3}\omega_{2,3}=\omega_{2,3}\\
\beta_{2,3}G^\hbar_{1,3}=G^\hbar_{1,2} & \beta_{2,3}\omega_{1,3}=\omega_{1,2}
\end{array}
\ee
\be
\label{relation-II}
(\beta_{1,2}\beta_{2,3})^3=\hbox{Id}.
\ee
Finally the quantum central element:
\bea
{\mathcal C}^\hbar&=&q^{-1/2}G^\hbar_{1,2}G^\hbar_{2,3}G^\hbar_{1,3}-
q^{-1}\bigl(G^\hbar_{1,2}\bigr)^2
-q \bigl(G^\hbar_{2,3}\bigr)^2-q^{-1}\bigl(G^\hbar_{1,3}\bigr)^2\nn\\
&&-q^{-1/2}\omega_{1,2}G^\hbar_{1,2}-q^{1/2}\omega_{2,3}G^\hbar_{2,3}-q^{-1/2}\omega_{1,3}G^\hbar_{1,3}
\label{M-el-q}
\eea
is chosen to be Hermitian: $\bigl({\mathcal C}^\hbar\bigr)^\dagger={\mathcal C}^\hbar$.

\section{Versal unfolding of the $D_4$ singularity.}

Given any  $\phi\in\mathbb C[u,v,w]$, the following formulae define a Poisson bracket on
$\mathbb C[u,v,w]$:
\be\label{eq:EG}
\{u,v\} =\frac{\partial\phi}{\partial w},\qquad \{v,w\}=\frac{\partial\phi}{\partial u},
\qquad \{w,u\}=\frac{\partial\phi}{\partial v},
\ee
and $\phi$ itself is a central element for this bracket, so that the quotient space
$$
M_\phi:= \mathbb C[u,v,w]\slash_{\langle\phi=0\rangle}
$$
inherits the Poisson algebra structure \cite{EG}.

For $\phi$ given by
\be\label{eq:delp}
\phi(u,v,w)= u^2+ v^2 + w^2- u v w ,
\ee
the quotient space $M_\phi$ has a simple $D_4$ singularity at the
origin. It was proved in \cite{EG} that all Poisson algebra
deformations of $(M_\phi,\{\cdot,\cdot\})$ are obtained by deforming
$\phi$ to:
$$
\tilde\phi= u^2+ v^2 + w^2- u v w + r_1 u + r_2 v + r_3 w+ r_4,
$$
where $r_1,r_2,r_3,r_4$ are any four complex parameters.
This means that on the deformed surface $M_{\widetilde\phi}=\mathbb
C[u,v,w]\slash_{\langle\widetilde\phi=0\rangle}$ the Poisson bracket
is still given by the formulae (\ref{eq:EG}) with $\phi$ substituted
by $\widetilde\phi$.

The equation $\widetilde\phi=0$ defines an affine irreducible cubic surface $M_{\widetilde\phi}$ in
$\mathbb C^3$ whose projective completion
$$
\overline M_{\widetilde\phi}:=\{(u,v,w,t)\in\mathbb P^3\,|
u^2 t+ v^2 t+ w^2 t- u v w + r_1 u t^2 + r_2 v t^2 + r_3 w t^2+ r_4 t^3=0\}
$$
is a del Pezzo surface of degree three and
differs from it by three smooth lines at infinity forming a triangle \cite{Obl}:
$$
t=0,\qquad u v w=0.
$$

Observe that this Poisson algebra  (\ref{eq:EG})  on
$M_{\widetilde\phi}$ coincides with our one (\ref{eq:gen-poisson1}),
(\ref{eq:gen-poisson2}) and (\ref{eq:gen-poisson3}), while
$\widetilde\phi$ coincides with the central element $\mathcal C$,
after the appropriate identifications:
$$
G_{12}\to u,\quad G_{13}\to v,\quad G_{23}\to w,\quad
\omega_{12}\to r_1,\quad \omega_{13}\to r_2,
\quad\omega_{23}\to r_3,
$$
and, thanks to  (\ref{eq:fr}),
$$
 -4+G_1G_2G_3G_\infty+G_1^2+G_2^2+G_3^2+G_\infty^2\to r_4.
$$
As a consequence our parametrisation (\ref{G12}) $G_{12}, G_{1,3}, G_{2,3}$ in terms of $Y_1,Y_2,Y_3$ defines an analytic surjective map
$$
\mu:\mathbb C^3\slash_{\langle Y_1+Y_2+Y_3={\rm const.}\rangle}\to M_{\widetilde\phi},
$$
giving rise to a system of flat coordinates for the  affine irreducible cubic surface
$M_{\widetilde\phi}$.

\begin{remark}
It is straightforward to prove that for $G_1=G_2=G_3=0$, the map $\mu$ is always invertible a part from the symplectic leaves for which
$$
Y_1+Y_2+Y_3 = i n \pi,\quad n\in\mathbb Z.
$$
In this case the Casimir element becomes
 $$
{ \mathcal C}=4-G_\infty^2= \left\{\begin{array}{lc}
 4&\hbox{for } n\hbox{ even,} \\
  0&\hbox{for } n\hbox{ odd,} \\ \end{array}
 \right.
 $$
and the symplectic leaves degenerate. In particular there exist two
points $(u,v,w)=(2,2,2)$ and $(u,v,w)=(0,0,0)$ for which each
$u,v,w$ are Casimirs, so that the symplectic leaves reduce to a
point.
\end{remark}

We stress that our quantisation procedure described in Section
\ref{se:quantum} is not only valid in the geometric case (i.e. when
we restrict this map to real non--negative $Y_1,Y_2,Y_3$) but can be
easily extended to all of  $\mathbb C^3\slash_{\langle
Y_1+Y_2+Y_3\rangle}$, thus providing an explicit and natural
quantisation of the affine cubic  surface $M_{\widetilde\phi}$.

We observe that in the case when each $u,v,w$ are Casimirs, Oblomkov
\cite{Obl} proved that  the quantisation of the affine cubic
surface $M_{\widetilde\phi}$ coincides with spectrum of the center
of the generalised rank $1$ double affine Hecke algebra studied in
\cite{Sa}.

\section{Poisson algebra structure on the monodromy data of the PVI equation}\label{se:pain}

In this section, we show that the Poisson algebra
(\ref{eq:gen-poisson1}),  (\ref{eq:gen-poisson2}) and
(\ref{eq:gen-poisson3}) is the image under the Riemann--Hilbert map
of the Lie--Poisson structure on
$\oplus_{1}^3\mathfrak{sl}(2,{\mathbb C})$. In order to do so, we
need to recall some well known facts about the Painlev\'e sixth
equation and its relation to the monodromy preserving deformations
equations (\cite{JMU,MJ1}).

\subsection{Isomonodromic deformations associated to the sixth Painlev\'e equation}\label{se:iso}

The Painlev\'e sixth equation PVI \cite{fuchs, Sch, Gar1},
\begin{eqnarray}
y_{tt}&=&{1\over2}\left({1\over y}+{1\over y-1}+{1\over y-t}\right) y_t^2 -
\left({1\over t}+{1\over t-1}+{1\over y-t}\right)y_t+\nn\\
&+&{y(y-1)(y-t)\over t^2(t-1)^2}\left[\alpha+\beta {t\over y^2}+
\gamma{t-1\over(y-1)^2}+\delta {t(t-1)\over(y-t)^2}\right],
\end{eqnarray}
is equivalent to the simplest non trivial case of the Schlesinger equations \cite{Sch}. These are Pfaffian differential equations
\begin{eqnarray}
&&
{\partial\over\partial u_j} {A}_i=
{[ {A}_i, {A}_j]\over u_i-u_j},\qquad i\neq j,\nn \\
&&
{\partial\over\partial u_i} {A}_i=
-\sum_{j\neq i}{[ {A}_i, {A}_j]\over u_i-u_j},\label{sch}
\end{eqnarray}
for $m\times m$ matrix valued functions ${A}_1=A_1(u),\dots,
{A}_{n}=A_n(u)$, $u=(u_1,\dots,u_n)$, where the independent variables $u_1$, \dots, $u_n$ are pairwise distinct.

The case corresponding to the PVI equation is for
$m=2$ and $n=3$ and describes  the monodromy preserving deformations
of a rank $2$ meromorphic connection over $\mathbb P^1$ with four
simple poles $u_1=0,u_2=1,u_3=t$, and $\infty$:
\begin{equation}\label{eq:fuchs}
\frac{{\rm d} \Phi}{{\rm d} \lambda} = \left(\frac{A_1(t)}{\lambda}+\frac{A_2(t)}{\lambda-t}+\frac{A_3(t)}{\lambda-1}
\right)\Phi,
\end{equation}
where
\begin{eqnarray}
&&{\rm eigen}(A_i)= \pm\frac{\theta_i}{2}, \quad\hbox{for } i=1,2,3,
\quad A_\infty:=-A_1-A_2-A_3\label{eq:eigen1}\\
&&
A_\infty =\left\{\begin{array}{l}
\left(\begin{array}{cc}\frac{\theta_\infty}{2}&\\
&-\frac{\theta_\infty}{2}\\
\end{array}\right),\qquad\hbox{for}\quad\theta_\infty\neq 0\\
\left(\begin{array}{cc}0&1\\
0&0\\
\end{array}\right),\qquad\hbox{for}\quad\theta_\infty= 0\\
\end{array}\right.\label{eq:eigen2}
\end{eqnarray}
and the parameters $\theta_i$, $i=1,2,3,\infty$ are related to the PVI parameters by
$$
\alpha=\frac{(\theta_\infty-1)^2}{2},\quad\beta=-\frac{\theta_1^2}{2},\quad \gamma=\frac{\theta_3^2}{2},
\quad \delta=\frac{1-\theta_2^2}{2}.
$$
The precise dependence of the matrices $A_1,A_2,A_3$ on the PVI solution $y(t)$ and its first derivative $y_t(t)$ can be found in \cite{MJ1}.

In this paper we take the monodromy matrices $M_1,M_2,M_3,M_\infty$
of the Fuchsian system (\ref{eq:fuchs}) defined w.r.t. the
fundamental matrix $\Phi_\infty$ normalized at $\infty$:
$$
\Phi_\infty = (\ID + {\mathcal O}\left(1/\lambda\right))\lambda^{-A_\infty}\lambda^{-R_\infty},
$$
where the term $\lambda^{-R_\infty}$ only appears in the resonant
case, i.e. when $\theta_\infty\in\mathbb Z_+$ in which case all
entries of $R_\infty$ are zero apart from $R_{\infty_{12}}$, or when
$\theta_\infty\in\mathbb Z_i$ in which case all entries of
$R_\infty$ are zero apart from $R_{\infty_{21}}$, and w.r.t. the
basis of loops  $l_1,l_2,l_3 $ with base point at $\infty$, where
$l_i$ encircles only once $u_i$, $i=1,2,3$, and  $l_1,l_2,l_3 $ are
oriented in such a way that
$$
M_1 M_2 M_3 M_\infty=\ID,
$$
where $M_\infty=\exp(2\pi i A_\infty)\exp(2\pi i R_\infty)$.

Denote by ${\mathcal F}(\theta_1,\theta_2,\theta_3,\theta_\infty)$
the moduli space of rank $2$ meromorphic connection over $\mathbb
P^1$ with four simple poles $0,1,t,\infty$ of the form
(\ref{eq:fuchs}) and by ${\mathcal
M}(\theta_1,\theta_2,\theta_3,\theta_\infty)$ the moduli space of
monodromy representations
$$
\rho:\pi_1(\mathbb P^1\setminus\{0,t,1,\infty\})\to SL_2({\mathbb C}))
$$
with prescribed local monodromies:
$$
{\rm eigen}(M_j)= \exp(\pm\pi i \theta_j),\quad j=1,2,3,\infty.
$$
Then the Riemann-Hilbert correspondence
$$
{\mathcal F}(\theta_1,\theta_2,\theta_3,\theta_\infty)\backslash{\mathcal G}\to
{\mathcal M}(\theta_1,\theta_2,\theta_3,\theta_\infty)\backslash GL_2({\mathbb C}),
$$
where $\mathcal G$ is the gauge group \cite{bol}, is defined by
associating to each Fuchsian system its monodromy representation
class.

\begin{theorem}\cite{hit}
The Riemann--Hilbert correspondence is a Poisson map.
\end{theorem}

Iwasaki \cite{iwa} proved that ${\mathcal M}(\theta_1,\theta_2,\theta_3,\theta_\infty)\backslash GL_2({\mathbb C})=M_{\widetilde\phi}$. We are now going to prove that  the Poisson
bracket (\ref{eq:EG}) on $M_{\widetilde\phi}$ is the image  under the Riemann-Hilbert map of the Poisson Lie bracket on  $\oplus_{1}^3\mathfrak{sl}^\ast(2,{\mathbb C})$.

\subsection{Poisson bracket on the monodromy data of the general Painlev\'e sixth equation}

The Schlesinger equations on $\mathfrak g:=\mathfrak{sl}(m,\mathbb C)$ admit Hamiltonian formulation
with time-dependent quadratic Hamiltonians
\begin{eqnarray}
&&
H_k= \sum_{l\neq k}
{{\rm Tr}\left({A}_k{A}_l\right)\over u_k-u_l},
\label{ham0}\\
&&
{\partial\over\partial u_k}{A}_l = \{{A}_l,H_k\},\label{ham000}
\end{eqnarray}
where $\{\cdot,\cdot\}$ is the standard Lie--Poisson bracket on  ${\mathfrak g}^*$, which can be represented in
$r$-matrix formalism:
$$
\left\{ {A}({\lambda_1})\otim_,{A}(\lambda_2)\right\}
= \left[ \one{A}({\lambda_1})+\two{A} (\lambda_2),{r}({\lambda_1} - \lambda_2)\right] ,
$$
where $r({z})=\frac{\Omega}{\lambda}$ is a {\it classical
$r$-matrix}, i.e a solution of the classical Yang--Baxter equation.
In the case of ${\mathfrak g}:=\oplus_{n} {\mathfrak{sl}}(m)$,
$\Omega$ is the {\it exchange matrix}\/ $\Omega=\sum_{i,j}\one{E_{ij}}\otim\two{E_{ji}}$ (we identify ${\mathfrak{sl}}(m)$ with its
dual by using the Killing form $(A,B)={\rm Tr}\, AB, \quad A, B \in
{\mathfrak{sl}}(m)$).

\noindent The standard Lie--Poisson bracket on
$\mathfrak{sl}(m,{\mathbb C})$ is mapped by the Riemann--Hilbert
correspondence to the {\it Korotkin--Samtleben bracket:}

\begin{eqnarray}\nn\label{eq:KSP}
&&
\left\{ M_i\otim_,M_i\right\}
=\frac{1}{2}\left(\two{M_i}\Omega\one{M_i} - \one{M_i} \Omega \two{M_i} \right)\\
&&
\\
&&
\left\{ M_i\otim_,M_j\right\}
=\frac{1}{2}\left(  \one{M_i} \Omega \two{M_j} +\two{M_j}\Omega\one{M_i}
-\Omega\one{M_i}\two{M_j}-\two{M_j}\one{M_i}\Omega\right) ,\quad\hbox{for}\,\,i<j.\nn
\end{eqnarray}
This bracket does not satisfy the Jacobi identity - however it
restricts to a Poisson bracket on the adjoint invariant objects.

\begin{theorem}\label{th:techKS}
In the PVI case the Korotkin--Samtleben
bracket restricted to the adjoint invariant objects
\be
G_{i,j}:=-\Tr (M_i M_j).
\label{def-Gij}
\ee
is given by the formulae (\ref{eq:gen-poisson1}), (\ref{eq:gen-poisson2}) and (\ref{eq:gen-poisson3}).
\end{theorem}

\begin{proof} We show how to prove relation (\ref{eq:gen-poisson1}), all the others being equivalent.
By definition of $G_{i,j}$ we have:
\begin{eqnarray}
\left\{G_{1,2},G_{2,3}\right\}&=&\{\Tr(M_1 M_2),\Tr(M_2 M_3)\}=
\onetwo\Tr\left(\left\{M_1\otim_, M_2\right\} \one M_2\two M_3+\right.\nn\\
&&+\two M_2\left\{ M_1\otim_, M_3\right\} \one M_2+
 \one M_1\left\{ M_2\otim_, M_2\right\}\two M_l+\nn\\
&&+\left.\one M_1\two M_2\left\{M_2\otim_, M_3\right\}
\right).\nn
\end{eqnarray}
Applying the Korotkin--Samtleben bracket (\ref{eq:KSP}), one gets:
\begin{eqnarray}
\left\{G_{1,2},G_{2,3}\right\}&=&\frac{1}{2} \onetwo\Tr
\left[\left( \one{M_1} \Omega \two{M_2} +\two{M_2}\Omega\one{M_1}-\Omega\one{M_1}\two{M_2}-
\two{M_2}\one{M_1}\Omega\right) \one M_2\two M_3+\right.\nn\\
&&+\two M_2\left( \one{M_1} \Omega \two{M_3} +\two{M_3}\Omega\one{M_1}-\Omega\one{M_1}\two{M_3}-
\two{M_3}\one{M_1}\Omega\right) \one M_2+\label{eq:pbproof}\\
&&+\one M_1\left(\two{M_2}\Omega\one{M_2}- \one{M_2} \Omega \two{M_2} \right) \two M_3+\nn\\
&&\left.+\one M_1\two M_2\left( \one{M_2} \Omega \two{M_3} +\two{M_3}\Omega\one{M_2}-\Omega\one{M_2}\two{M_3}-
\two{M_3}\one{M_2}\Omega\right)\right] .
\nn
\end{eqnarray}
In the subsequent calculations we use that $\Omega$ is the exchange matrix,
which implies that for every $i, j$:
\begin{equation}
\two{M_j}\Omega\one{M_i}=\Omega\one{M_j}\one{M_i}=\two{M_j}\two{M_i}\Omega.
\label{one-two}
\end{equation}
We then obtain that the
first two lines on the right hand side of (\ref{eq:pbproof}) cancel
each other and:
$$
\left\{G_{1,2},G_{2,3}\right\}= \Tr\left(M_1 M_2 M_3 M_2 -
M_1 M_2 M_2 M_3
\right).
$$
By repeated applications of the {\it skein relation:}
\begin{equation}\label{eq:skein}
\Tr( A B) + \Tr (A B^{-1}) = \Tr (A) \Tr(B),
\end{equation}
which is valid for any $2\times2$-matrices $A$ and $B$ with unit
determinants, we obtain the final result. In fact
\begin{eqnarray}
&&
 \Tr\left(M_1 M_2 M_3 M_2\right) = \Tr\left( M_1M_2 \right) \Tr\left( M_2 M_3 \right)+
  \Tr\left( M_1 M_3\right)- \Tr(M_1)\Tr(M_3)
  \nn\\
&&  \Tr\left(M_1 M_2 M_2 M_3\right) = \Tr\left( M_1M_2 \right) \Tr\left( M_2 M_3 \right)-\Tr(M_1)\Tr(M_3)
+\nn\\
&&
\quad\qquad\qquad\qquad\qquad+\Tr(M_\infty) \Tr(M_2)-  \Tr\left( M_1 M_2  M_3 M_2\right),\nn
  \end{eqnarray}
  so that
  $$
  \left\{G_{1,2},G_{2,3}\right\}= -2 G_{1,3} + G_{1,2} G_{2,3}-\omega_{1,3}
  $$
  as we wanted to prove.  The other relations can be obtained in a similar way. The Jacobi identity is a straightforward brute force computation.
\end{proof}

\begin{remark}
Observe that Dubrovin produced the following Poisson bracket on the Stokes data associated to a $3$-dimensional Frobenius manifold:
\begin{eqnarray}
&&
\left\{S_{1,2},S_{2,3}\right\}= S_{1,2} S_{2,3} -2 S_{1,3} ,\label{eq:dub-poisson1}\\
&&
\left\{S_{2,3},S_{1,3}\right\}= S_{2,3} S_{1,3} -2 S_{1,2} ,\label{eq:dub-poisson2}\\
&&
\left\{S_{1,3},S_{1,2}\right\}=  S_{1,2} S_{1,3} -2 S_{2,3}.\label{eq:dub-poisson3}
\end{eqnarray}
It is a straightforward computation to show that this bracket
coincides with our bracket for the case of PVI$\mu$, i.e. the
Painlev\'e sixth equation with parameters $\beta=\gamma=0$,
$\delta=\frac{1}{2}$ and $\alpha=\frac{2\mu-1}{2}$ appearing in the
Frobenius manifold theory,  via the change of coordinates
$$
G_{i,j} = S_{ij}^2-2.
$$ 
This change of coordinates actually corresponds to a quartic transformation on the sixth Painlev\'e equation \cite{MV}.
\end{remark}

\subsection{Action of the braid group $B_3$}

The procedure of the analytic continuation of the solutions of the
PVI equation was described in  \cite{DM} by  the action of the braid
group ${\mathcal B}_3=\langle \beta_{12},\beta_{23}\rangle$ on the
monodromy matrices $M_1,M_2,M_3$ given by
\begin{eqnarray}\label{eq:braidDM}
&&
\beta_{12}(M_1,M_2,M_3)=(M_1M_2 M_1^{-1} ,M_1,M_3)\\
&&
\beta_{23}(M_1,M_2,M_3)=(M_1,M_2 M_3 M_2^{-1},M_2).
\end{eqnarray}
By using the skein relation (\ref{eq:skein}) it is a straightforward
computation to prove that the action of the braid group on the
Poisson algebra  (\ref{eq:gen-poisson1}), (\ref{eq:gen-poisson2}),
and (\ref{eq:gen-poisson3}) is given by formulae (\ref{beta1}) and
(\ref{beta2}).

\begin{remark}
In the Teichm\"uller space framework all coordinates $Y_1,Y_2,Y_3$
are assumed to be positive real numbers, therefore the products
$\gamma_i\gamma_j$ are always hyperbolic elements, i.e.
$G_{ij}=-\Tr(\gamma_i\gamma_j)>2$.  Therefore there is no
Teichm\"uller space interpretation of the algebraic solutions of PVI
\cite{DM}.
\end{remark}


\end{document}